# Phase-shifting technique for improving the imaging capacity of sparse-aperture optical interferometers

François Hénault

*UMR 6525 H. Fizeau, Université de Nice-Sophia Antipolis*

*Centre National de la Recherche Scientifique, Observatoire de la Côte d'Azur*

*Avenue Nicolas Copernic, 06130 Grasse – France*

In this paper, we describe the principle of a multi-aperture interferometer that uses a phase-shifting technique and is suitable for quick, snapshot imaging of astrophysical objects at extreme angular resolution through Fourier inversion. A few advantages of the proposed design are highlighted, among which are radiometric efficiency, Field of View equivalent to those of Fizeau interferometers, and a preliminary calibration procedure allowing characterization of instrumental errors. For large telescope numbers, the proposed design also results in considerable simplification of the optical and mechanical design. Numerical simulations suggest that it should be possible to couple hundreds of telescopes on a single 4K x 4K detector array, using only conventional optical components or emerging technologies.

## 1  INTRODUCTION

Long baseline interferometry in the infrared (IR) or optical domain is a very powerful tool for extremely high angular resolution imaging of astronomical objects. It was originally inspired by the brilliant successes of radio astronomy and based on Fourier inversion of interferometric fringe parameters [1-2]. But it is only very recently that detailed images of nearby stars can be obtained in this particular wavelength range [3-8], because it requires a coherent combination of the beams collected by all the telescopes constituting the interferometer, with an equalization of their different Optical Path Differences (OPD) typically lower than one micrometer: nowadays this major technical issue can be considered as solved, and modern installations such as the Very Large Telescope Interferometer (VLTI) or Center for High Angular Resolution Astronomy (CHARA) aim at providing in the forthcoming years a vast amount of visible and IR science images with the help from new generation combining instruments such as VEGA [9] or the Michigan Infrared Combiner (MIRC) on CHARA [10] – the latter instrument can for example combine up to 6 telescopes and obtain 15 squared visibilities and 10 independent closure phases in quasi-real time. However it has to be noticed that most of the produced images result from a long data processing work, on the one hand, and that the total number of collecting telescopes in this waveband still remains lower than in radio-interferometric facilities such as the Very Large Array (VLA) in New Mexico.

In parallel, new concepts for "direct imaging" interferometers were proposed during the last decades: just like a conventional telescope, a direct imaging system aims at providing well-resolved, "snapshot" pictures of the observed sky object in a reasonable lapse of time. The most famous of these concepts actually is the Fizeau interferometer, a multi-aperture optical system made of a number of separated telescopes obeying to a "golden rule" stating that the output pupil of the whole system should be a scaled replica of its entrance pupil [11-12]. But it must be noticed that the largest Fizeau interferometer ever built currently is the Large Binocular Telescope (LBT) [13] whose 22-m baseline remains quite modest. In addition the optical and mechanical architecture of a Fizeau interferometer is considered as much more difficult, because many optical components need to be manufactured, aligned and actively controlled in order to relay the image and pupil planes from the collecting telescopes to the final combiner, and these optics should be packed on a very restricted volume. Practically, their arrangement might become impracticable above a given number N of telescopes (say, N ≥ 10). A good illustration of such complexity can be found in the paper from Meinel and Meinel [14] who described the opto-mechanical design of a space telescope made of numerous phased sub-apertures.

A few variants of the Fizeau interferometer have already been proposed in order to improve the Signal to Noise Ratio (SNR) at the expense of a reduced Field of View (FoV) [15-17], and some other get rid of the opto-mechanical complexity by replacing most of the relay optics with monomode optical fibers [18]. But herein one of our main goals is to define an alternative optical design suitable for quick or "snapshot" imaging, applicable with any number N of telescopes, and that should stay independent of any technological solution (this point is further discussed in section 5.2). For that purpose it is makes use of a phase-shifting technique originally intended for focal plane wavefront sensing [19-20], associated with optical remapping of the output pupil: in sections 2 and 3 are respectively described a preliminary optical design of the whole system and the theoretical principles of the method. Results of numerical simulations validating the proposed concept are presented in section 4. Short discussions about practical implementation, effective FoV, SNR, array redundancy and a tentative calibration procedure are provided in section 5. Finally, the major conclusions are summarized in section 6.

## 2   INSTRUMENTAL CONCEPT

The instrumental design proposed here is more especially intended for space applications. It is principally organized as an array of free-flying telescopes and a central combining spacecraft, being sketched in Figure 1 to Figure 3. The collecting telescopes are positioned on a giant, fictitious parabolic or spherical surface whose focus is localized at the entrance of the beam-combining unit. It is assumed that the whole telescope array is slightly decentered with respect to the focal point of the fictitious giant surface so that the central combiner spacecraft does not vignet any telescope beam. Spacecraft repositioning movements needed to point off-axis targets could be minimized by combining transverse displacements of the central vessel with axial shifts of the collecting telescopes compensating for spherical aberration (see Figure 1), or alternatively their internal Optical Delay Lines (ODL) might be utilized for that purpose.

All collecting vessels are supposed to be identical, each comprising an afocal Mersenne telescope and an ODL of the cat's eye type, compensating for OPD errors and relaying the telescope pupil (assumed to be its primary mirror) towards the combining optics as depicted in Figure 2. A Tip-Tilt Mirror (TTM) finally directs the parallel beam in the direction of the central combiner. The joint use of ODLs and TTMs ensures an easy reconfiguration of the interferometric array and thus a high flexibility of the whole system.

The beam combining optics are schematically represented in Figure 3. From their entrance to the focal plane array (assumed to be a conventional detector matrix), they are made of three major optical subsystems:

1) A matrix of tip-tilt mirrors in charge of collecting the beams originating from each telescope and reflecting them along the Z-axis direction.

2) An optical subsystem provisionally denominated "pupil remapping optics" whose possible designs will be discussed in section 5.2. For now let us just consider that its main optical functions are the following: a) – To perform a geometrical remapping of the input pupil into a series of output sub-pupils centered on equally-spaced points along an axis denoted X'. b) – To preserve imaging between the $n^{th}$ entrance and exit sub-pupils ($1 \leq n \leq N$) with sufficient image quality. c) – To coarsely preserve the OPD equalization betweens the different interferometer arms, knowing that the individual ODLs of the collecting telescope can be used for fine adjustment. It has to be noticed that this optical subsystem deliberately violates the golden rule of Fizeau interferometers, and as a consequence no direct images of the object should be observable in the image plane.

3) Anamorphic focusing and dispersive optics, schematically represented by a prism and a couple of cylindrical lenses in Figure 3, although a design based on two cylindrical mirrors and a diffraction grating might be efficient as well. Anamorphic elements are used to concentrate the interference pattern generated by the output pupil near the X"-axis of the Figure, while the dispersive component (whose axis is assumed to be perpendicular) allows recording simultaneously monochromatic fringes at different wavelengths $\lambda$ on the same detector chip.

It must be emphasized again that instead of images, one will observe a set of fringes perpendicular to the X"-axis, carrying all the information necessary to reconstruct the sky object when the interferometer is operated in phase-shifting mode. The purpose of the next section is then to present the theory of such a "phase-shifting stellar interferometer" instrument.

## 3  THEORY

The theory behind this study essentially derives from the formalism previously published in Refs. [20] and [21], however it is fully summarized here in order to make this paper self-consistent. Additionally, a few simplifications of the mathematical notation later introduced in Ref. [22] are also taken into account.

### 3.1  Scientific notations and basic relationships

The four employed coordinate systems are represented in Figure 4. They all share one common, principal optical axis Z directed at the FoV centre. From distant sky to the final detector plane, these coordinate systems are the following:

- The ($\alpha,\delta$) axes are defining an on-sky reference system allowing to localize any sky object by angular coordinates ($\alpha,\delta$) standing for example for right ascension and declination. The object direction can also be defined by a unit vector **s** (bold characters denoting vectors throughout the whole paper) of direction cosines $\mathbf{s} \approx (1,\alpha,\delta)$ if $\alpha$ and $\delta$ are considered as first-order quantities.

- All the collecting telescopes are assumed to lie in a common input pupil plane (O,X,Y) where O is a reference point located near the array centre. In this coordinate system the sub-aperture centres are defined by points $P_n$, with index n comprised between 1 and N (N being the total number of collecting telescopes).

- In the same way, all the output sub-pupils relayed from the telescopes lay in a common exit pupil plane (O',X',Y'), and their centres are located at points $P'_n$.

- Finally, the coordinate system (O",X",Y") is attached to the focal plane of the multi-aperture interferometer, although this reference frame does not appear explicitly in the presented formalism.

Let us now define the main hypotheses in which the general theory is applicable:

1) All the collecting telescopes and their optical train conveying the beam to the combining optics are assumed to have identical entrance pupil diameters D and output pupil diameters D'.

2) For $1 \leq n \leq N$, the $n^{th}$ collecting aperture centered on $P_n$ is optically conjugated with its associated output sub-aperture centered on $P'_n$ without geometrical aberrations.

3) All the theory is only valid in the frame of first-order Gaussian optics and Fraunhofer scalar diffraction.

From the above hypotheses, it was shown in Refs. [21-22] that the irradiance distribution I(**s**) of the image formed by the multi-aperture interferometer and projected back onto the sky

(thus avoiding the use of the image plane reference frame) can be written in a first-order approximation:

$$I(\mathbf{s}) = \iint_{\mathbf{r} \in \Omega_O} O(\mathbf{r}) \left| \sum_{n=1}^{N} \hat{B}_D(\mathbf{s}-\mathbf{r}) \, a_n \exp[i\varphi_n] \exp[ik(\mathbf{r\,OP_n} - \mathbf{s\,O'P'_n}/m)] \right|^2 d\mathbf{r}, \qquad (1)$$

with the following definitions:

- **s** and **r** both stand for a unit vector directed at any point in the sky (and corresponding to any point M" or P" in the image plane, see Figure 4),

- O(**s**) is the angular brightness distribution of the extended sky object that is being observed by the interferometer,

- $\Omega_O$ is the solid angle subtended by the observed FoV including all vectors **s**,

- $\hat{B}_D(\mathbf{s})$ is the complex amplitude created at the focal plane of an individual collecting telescope and back-projected onto the sky. For an unobstructed circular pupil it is equal to $2J_1(\rho)/\rho$, where $\rho = k\,D\,\|\mathbf{s}\|/2$ and $J_1$ is the type-J Bessel function at the first order,

- $k = 2\pi/\lambda$ is the wave number of the electro-magnetic field assumed to be monochromatic, and $\lambda$ is its wavelength in vacuum,

- $a_n$ and $\varphi_n$ respectively are the amplitude transmission factor and phase-shift introduced along the $n^{th}$ interferometer arm,

- $m = D'/D$ is the optical compression factor of the system.

As mentioned in Ref. [21], Eq. (1) is a generalized Object-Image relationship that can only be simplified if certain conditions are fulfilled: one of these precisely applies to the Fizeau interferometer case, because its intrinsic "golden rule" can be written as:

$$\mathbf{O'P'_n} = m\,\mathbf{OP_n} \qquad (2)$$

for all indices n comprised between 1 and N (the respect of this condition ensures that both the diameters and geometry of the input and output sub-apertures are homothetic). Combining together Eqs. 1 and 2 allows retrieving the classical Object-Image relationship of Fourier optics:

$$I(\mathbf{s}) = PSF(\mathbf{s}) * O(\mathbf{s}),  \tag{3}$$

where symbol * stands for the convolution product and PSF(**s**) is the Point Spread Function (PSF) of the complete optical system that is:

$$PSF(\mathbf{s}) = \left|\hat{B}_D(\mathbf{s})\right|^2 \left|\sum_{n=1}^{N} a_n \exp[i\varphi_n]\exp[i\,k\,\mathbf{s}\,\mathbf{OP_n}]\right|^2. \tag{4}$$

Let us now return to the most general case where the input and output sub-pupils of the interferometer have independent geometries, as in a vast majority of existing facilities.

## 3.2 Addition of a phase-shifted reference pupil

As sketched in Figure 4, the basic principle of the method consists in adding a reference sub-aperture in the entrance pupil plane (O,X,Y), and to record the fringes formed at the focal plane of the interferometer when successive phase-shifts $\phi_m$ are applied into the reference pupil. Combining different phase-shifted interferograms then allows estimating the amplitude and phase of the Mutual Coherence Function (MCF), which by virtue of Van Cittert-Zernike theorem [23] is the Fourier transform of the object brightness distribution O(**s**). It is assumed that the reference pupil is located at the centre O of the telescope array and has the same diameter D than the other sub-pupils. We also suppose the interferometer to be perfectly phased (i.e. $\varphi_n = 0$ for all sub-apertures excepted the central one where the reference phase-shift will be introduced) and having identical amplitude transmission factors $a_n$ that can be set arbitrarily to 1. Hence the expression of the phase-shifted intensity distribution $I_m(\mathbf{s})$ can be derived from Eq. 1:

$$I_m(\mathbf{s}) = \iint_{\mathbf{s_o} \in \Omega_O} O(\mathbf{r}) \left| \hat{B}_D(\mathbf{s}-\mathbf{r}) \exp[i\phi_m] + \sum_{n=1}^{N} \hat{B}_D(\mathbf{s}-\mathbf{r}) \exp[ik\zeta_n(\mathbf{r},\mathbf{s})] \right|^2 d\mathbf{r}, \quad (5)$$

where $\zeta_n(\mathbf{r},\mathbf{s})$ is a condensed notation for the sum of external and internal OPDs:

$$\zeta_n(\mathbf{r},\mathbf{s}) = \mathbf{r}\,\mathbf{OP}_n - \mathbf{s}\,\mathbf{O'P'}_n / m. \quad (6)$$

Expanding the square modulus of the amplitude function inside the integral in Eq. (5) leads to a mathematical expression composed of four terms:

$$I_m(\mathbf{s}) = \iint_{\mathbf{s_o} \in \Omega_O} O(\mathbf{r}) \left| \hat{B}_D(\mathbf{s}-\mathbf{r}) \right|^2 d\mathbf{r}$$

$$+ \iint_{\mathbf{r} \in \Omega_O} O(\mathbf{r}) \left| \sum_{n=1}^{N} \hat{B}_D(\mathbf{s}-\mathbf{r}) \exp[ik\zeta_n(\mathbf{r},\mathbf{s})] \right|^2 d\mathbf{r}$$

$$+ \exp[i\phi_m] \iint_{\mathbf{r} \in \Omega_O} O(\mathbf{r}) \left| \hat{B}_D(\mathbf{s}-\mathbf{r}) \right|^2 \sum_{n=1}^{N} \exp[-ik\zeta_n(\mathbf{r},\mathbf{s})] d\mathbf{r}$$

$$+ \exp[-i\phi_m] \iint_{\mathbf{r} \in \Omega_O} O(\mathbf{r}) \left| \hat{B}_D(\mathbf{s}-\mathbf{r}) \right|^2 \sum_{n=1}^{N} \exp[ik\zeta_n(\mathbf{r},\mathbf{s})] d\mathbf{r}. \quad (7)$$

Denoting $PSF_D(\mathbf{s}) = \left| \hat{B}_D(\mathbf{s}) \right|^2$ the Point Spread Function of an individual aperture being projected back onto the sky, the previous relationship may be rewritten as:

$$I_m(\mathbf{s}) = I_O(\mathbf{s}) + O(\mathbf{s}) * \text{PSF}_D(\mathbf{s}) + \exp[i\phi_m] \iint_{\mathbf{r} \in \Omega_O} O(\mathbf{r}) \, \text{PSF}_D(\mathbf{s}-\mathbf{r}) \sum_{n=1}^{N} \exp[-ik\zeta_n(\mathbf{r},\mathbf{s})] \, d\mathbf{r}$$

$$+ \exp[-i\phi_m] \iint_{\mathbf{r} \in \Omega_O} O(\mathbf{r}) \, \text{PSF}_D(\mathbf{s}-\mathbf{r}) \sum_{n=1}^{N} \exp[ik\zeta_n(\mathbf{r},\mathbf{s})] \, d\mathbf{r}, \qquad (8)$$

where the expression of $I_O(\mathbf{s})$ is rigorously similar to Eq. 1 in the case when $\varphi_n = 0$ and $a_n = 1$ (i.e. the interferometer is phased). The heart of the method now consists in combining digitally M different phase-shifted fringes $I_m(\mathbf{s})$ using a specific set of $\phi_m$ values satisfying both conditions [20]:

$$\sum_{m=1}^{M} \exp[i\phi_m] = \sum_{m=1}^{M} \exp[2i\phi_m] = 0. \qquad (9)$$

Then a linear combination $I_C(\mathbf{s})$ of the M functions $I_m(\mathbf{s})$ with the complex coefficients $\exp[i\phi_m]$ allows canceling all the terms of Eq. (8) excepting the fourth and last one:

$$I_C(\mathbf{s}) = \frac{1}{M} \sum_{m=1}^{M} \exp[i\phi_m] I_m(\mathbf{s}) = \iint_{\mathbf{r} \in \Omega_O} O(\mathbf{r}) \, \text{PSF}_D(\mathbf{s}-\mathbf{r}) \sum_{n=1}^{N} \exp[ik\zeta_n(\mathbf{r},\mathbf{s})] \, d\mathbf{r}. \qquad (10)$$

The way of selecting an appropriate set of phase-shifts $\phi_m$ is quite straightforward and has been explained in Ref. [20]. The present study is voluntarily restricted to the most simple set of possible phase steps $\phi_m$, where M = 3 and $\phi_1 = 0$, $\phi_2 = 2\pi/3$ and $\phi_3 = 4\pi/3$. Inserting again expression (6) in Eq. 10 and expanding the complex exponential out of the integral subsequently leads to an alternative expression of $I_C(\mathbf{s})$:

$$I_C(\mathbf{s}) = \sum_{n=1}^{N} \exp[-ik\mathbf{s}\mathbf{O'P'}_n/m] \iint_{\mathbf{r} \in \Omega_O} O(\mathbf{r}) \, \text{PSF}_D(\mathbf{s}-\mathbf{r}) \exp[ik\mathbf{r}\mathbf{OP}_n] \, d\mathbf{r}$$

$$= \sum_{n=1}^{N} \exp[-ik\mathbf{s}\mathbf{O'P'}_n/m] \left\{ O(\mathbf{s}) \exp[ik\mathbf{s}\mathbf{OP}_n] * \text{PSF}_D(\mathbf{s}) \right\}. \qquad (11)$$

The next step is then to return into the entrance pupil plane (OXY) that is closely related to the Optical Transfer Function (OTF) plane of angular frequencies (the "u-v plane" as often named by specialists). Inverse Fourier transforming of Eq. (11) readily allows computing the function $OTF_C(P)$, which may be regarded as the OTF of the entire phase-shifted array. After elementary manipulations of Fourier transform theorems, one finds a rather simple expression of $OTF_C(P)$:

$$OTF_C(P) = FT^{-1}[I_C(s)] = \sum_{n=1}^{N} \delta(P - P'_n/m) * \{\hat{O}(P + P_n) \, OTF_D(P)\}$$

$$= \sum_{n=1}^{N} \hat{O}(P - P'_n/m + P_n) \, OTF_D(P - P'_n/m), \qquad (12)$$

where $\delta(P)$ is the impulse Dirac distribution, function $\hat{O}(P)$ is the inverse Fourier transform of the object brightness distribution (i.e. its complex MCF), and $OTF_D(P)$ simply stands for the OTF of an individual telescope aperture of diameter D. Multiplying both sides of Eq. 12 by $\delta(P - P'_n/m)$ finally provides an estimation of the complex MCF on the $n^{th}$ sub-pupil of the interferometer:

$$\delta(P - P'_n/m) \, OTF_C(P) = \hat{O}(P_n) \, OTF_D(0) = \hat{O}(P_n), \qquad (13)$$

since $OTF_C(0) = 1$ by definition. The last step of this procedure consists in invoking the Van Cittert-Zernike theorem [23] in order to reconstruct the original sky object by means of a last, discrete inverse Fourier transform. Hence, remembering that $k = 2\pi/\lambda$:

$$O_R(s/\lambda) = \sum_{n=1}^{N} \hat{o}_n \exp[i\theta_n] \exp[2i\pi s OP_n/\lambda] \qquad (14)$$

where $\hat{o}_n$ and $\theta_n$ respectively are the modulus and phase of $\hat{O}(P_n)$, both being extracted from $OTF_C(P)$ via the simple relationships:

$$\hat{o}_n = \left| \delta(P - P'_n/m) \, OTF_C(P) \right|$$

$$\theta_n = Arg\{\delta(P - P'_n/m) \, OTF_C(P)\} \quad Mod[2\pi]. \tag{15}$$

It might be objected that the here above method is not a true, direct imaging process since it makes use of a Fourier transform reconstruction algorithm instead of catching snapshot images on a detector array. But practically the whole computing time required by Eqs. (10) to (15) is very short and suitable to quasi-real time applications, as will be confirmed by the numerical simulations presented in section 4.

## 4   NUMERICAL SIMULATIONS

In this section are provided the numerical results of computer simulations that were undertaken to assess the imaging capability of an interferometer having linearly remapped output sub-pupils: we consider a multi-aperture interferometer whose main optical characteristics are summarized in Table 1. The array consists of 49 collecting telescopes (a number inspired by the current development of the Atacama Large Millimeter/sub millimeter Array (ALMA) [24]), geometrically arranged on a square grid as depicted on the upper panel of Figure 5. Here the combining optics are not described in detail, but it is assumed that their general scheme is the same as in section 2, Figure 3. All numerical computations are monochromatic and are carried out at the thermal IR wavelength $\lambda = 10$ µm, which is one of the most favorable options for the characterization of extra-solar systems and of their forming disks and planets.

Table 1: Optical and geometrical characteristics of simulated interferometers.

| Optical/geometrical parameters | Linearly remapped output pupil interferometer (§ 4) | Fizeau interferometer (§ 5.1) |
| --- | --- | --- |
| Number of collecting telescopes N | 49 | 49 |
| Telescope diameter D | 1 m | 1 m |
| Telescope focal length F | 50 m | 50 m |
| Maximal entrance baseline B | 100 m | 100 m |
| Focal length of relay optics $F_C$ | 100 mm | 100 mm |
| Focal length of combining optics F' | 1 m | 1 m |
| Output sub-pupil diameter D' | 2 mm | 2 mm |
| Maximal output baseline B' | 275 mm | 200 mm |
| Pixels number $n_P$ | 729 x 75 | 729 x 729 |
| Pixel size | 9 x 9 µm | 9 x 9 µm |

All the numerical simulations follow the rationale indicated on the left side of Figure 6. The computer program is basically split into an image simulation procedure and an object reconstruction algorithm as shown on the top and bottom halves of the Figure. The major steps of the program are now detailed below.

1) First the angular brightness distribution O(**s**) of the sky object must be defined. In this paper are considered three different fictitious sky-objects labelled O#1, O#2 and O#3. O#1 is a simple uniform disk of approximately 40 mas diameter (not shown on the Figures), O#2 looks like a dust ring featuring a decentered star (top left panel of Figure 7) and O#3 is an extended object surrounded by a thin ring (top left panel of Figure 8). Functions O(**s**) are imported into the program from 45 x 45 external files encoded in 256 grey levels, whose square sides correspond to 80 milli-arcseconds

(mas) on-sky. The object file is embedded into a 729 x 729 array filled with zeros before further processing.

2) Input and output pupil maps of all the sub-pupils are computed. The input pupil mainly serves for verification and illustration purposes (such as in Figure 5) while the output pupil is the reference map for amplitude and phase extraction (see step 7). Here the sub-pupils are realigned along the X'-axis and packed into a very compact and redundant arrangement where the gap between two neighboring sub-apertures is roughly equal to their half-diameter. The output pupil mask is stored into a 729 x 75 rectangular array, as well as all the fringe maps computed later.

3) A loop is then started for all the selected phase-shifts $\phi_1 = 0$, $\phi_2 = 2\pi/3$ and $\phi_3 = 4\pi/3$ (many other combinations involving more phase steps are possible [20], but this is the simplest and most computationally rapid).

4) For each $\phi_m$, a direct numerical integration of the integral in Eq. (5) is carried out: this is obviously the most demanding computing effort of the program and may take several hours of computing time.

5) Once all fringe patterns $I_m(\mathbf{s})$ are computed, the loop is closed and their linear combination $I_C(\mathbf{s})$ is calculated according to Eq. 10: this is the beginning of the object reconstruction process.

6) The inverse Fourier transform of $I_C(\mathbf{s})$ is computed via a classical FFT algorithm in application of Eq. 12. The resulting function $OTF_C(P)$ is stored into a bi-dimensional complex array.

7) For $1 \leq n \leq N$, the amplitudes $\hat{o}_n$ and phases $\theta_n$ of the mutual coherence function $\hat{O}(P_n)$ can now be extracted on all the sub-apertures of the array (Eqs. 15). Referring to the output pupil map prepared in step 2, $\hat{o}_n$ is taken as the maximal value of $|OTF_C(P)|$ found on linear segments of the $n^{th}$ output sub-pupil along the X'-axis, and

$\theta_n$ as the phase of $OTF_C(P)$ at that point (it has been checked that averaging the phases on the full sub-pupil leads to similar numerical results).

8) Lastly, the complex mutual coherence function $\hat{O}(P_n)$ is built from the evaluated amplitudes and phases $\hat{o}_n$ and $\theta_n$ in a 45 x 45 array previously initialized to zero (excepted for the null angular frequency at the array centre where by definition $\hat{o}_0 = 1$ and $\theta_0 = 0$). Fourier transforming of this array according to Eq. (14) thus allows reconstructing the observed sky object with the same original angular scale. This reconstructed brightness distribution is noted $O_R(\mathbf{s})$.

Figure 7 shows an illustrative example of the whole process, from the acquisition of phase-shifted fringes to final object reconstruction. Here the considered sky object O#2 is shown on Figure 7a. In Figures 7b and c are presented the images that would be formed by a monolithic telescope of 5 or 1-meter diameter for comparison purpose. A typical fringe pattern $I_1(\mathbf{s})$ generated by the linearly remapped output pupil interferometer is depicted in Figure 7d. Figures 7e and f respectively show the next recorded interferogram $I_2(\mathbf{s})$ that is $2\pi/3$ phase-shifted with respect to $I_1(\mathbf{s})$, and the difference between both previous interferograms in order to highlight their intensity variations (since the fringes look identical at first glance). The maximal difference is typically found around 10 % of the maximal fringe intensity.

Figures 7g and h exhibit grey-scaled maps of the modulus and phase of function $OTF_C(P)$ calculated from Eqs. (10) and (12) and looking as a series of parallel lines, each being associated to a given sub-aperture. A three-dimensional (3D) view of the modulus of $OTF_C(P)$ is also shown in Figure 7i, where the peak heights are proportional to the amplitudes $\hat{o}_n$ of the mutual coherence function $\hat{O}(P)$. A three-dimensional view of the MCF modulus is reproduced on Figure 7j. Figures 7k-m finally display 3D views of the modulus $|O_R(\mathbf{s})|$ and imaginary part $Im[O_R(\mathbf{s})]$ of the reconstructed sky object, as well as a grey scale map to be compared with the original object on the top left panel. Here the quality of the full image reconstruction process may be estimated by means of the imaginary part that should ideally be uniformly equal to zero since by definition the angular brightness distribution of the observed object is a real (and positive) number. For all the performed numerical simulations the characteristics of $Im[O_R(s)]$ are summarized in Table 2, where they are expressed in terms of PTV and RMS percentage of the maximal value of $|O_R(s)|$: it can be noticed that they are all around 1 and 0.2 % respectively.

Table 2: Imaginary part of reconstructed sky objects $O_R(s)$ expressed in terms of PTV and RMS percentage of the maximal modulus.

| Sky maps | Peak to Valley (PTV) | Root Mean Square (RMS) | 2D or 3D plots |
|---|---|---|---|
| O#1, linearly remapped output pupil interferometer | 0.87 % | 0.17 % | Not shown |
| O#2, linearly remapped output pupil interferometer | 1.20 % | 0.28 % | Figure 7l |
| O#3, linearly remapped output pupil interferometer | 0.73 % | 0.16 % | Not shown |
| O#1, Fizeau interferometer | 0.04 % | 0.01 % | Not shown |
| O#2, Fizeau interferometer | 0.59 % | 0.14 % | Not shown |
| O#3, Fizeau interferometer | 0.18 % | 0.04 % | Not shown |

Obviously the fidelity of the reconstructed objects could be improved by using some of the classical image enhancement algorithms introduced in radio and optical interferometry [1-2]: one may think for example about the CLEAN algorithm [25] as an ideal companion of this type of interferometer. However the main purpose of this paper is not to optimize the object reconstruction process, but more modestly to propose a simplified opto-mechanical design for combining a large number of separate telescopes. Here a remarkable point resides in the rapidity of image reconstruction (i.e. steps 5-8 of the general procedure) that only takes a few seconds of computing time in all the studied cases. For this reason, a phase-shifting, linearly remapped output pupil interferometric array having a sufficient number of telescopes may finally be also considered as a snapshot imaging instrument, without suffering from the intrinsic complexity of Fizeau interferometers, at least when the required integration time on the detector does not exceed the processing time.

## 5 DISCUSSION

The present section deals with some theoretical and practical considerations about the real imaging power of the proposed interferometer design and the way it could effectively be built using only existing technologies. Below are discussed different topics such as achievable Field of View, SNR and chromatism (§ 5.1), possible techniques for realizing the critical pupil remapping optics subsystem (§ 5.2), and a tentative calibration procedure suitable for space applications (§ 5.3).

### 5.1 Theoretical considerations

*Effective Field of View*

The effective FoV of a multi-aperture interferometer is a tricky subject that has been discussed by many authors over the years. It is commonly admitted that a necessary condition for obtaining a wide imaging FoV is to respect the "golden rule of Fizeau interferometers" stating that their output pupil must be a strict homothetic replica of their entrance pupil [11-12], and herein expressed by Eq. 2. However the latter condition is not sufficient since it is only relevant in the frame of first-order optics: higher-order aberrations such as linear or quadratic piston errors may also limit the total FoV of the interferometer, as discussed comprehensively in Ref. [26].

In order to compare the effective FoV of the remapped output pupil interferometer with the theoretically infinite FoV of the Fizeau interferometer (at least at first order), additional numerical simulations were carried out for the latter case, using a slightly modified version of the computer program (as sketched on the right side of Figure 6) and updated geometrical parameters (right column of Table 1): here the major change is the total number of pixels $n_P$, being now equal to 729 x 729 in the Fizeau case. The main results are provided in Figure 8, showing the original sky object O#3 (Figure 8a), its image formed by the Fizeau interferometer (Figure 8b), a grey scale map of the reconstructed images (Figure 8c), and a 3D view of their difference map (Figure 8d). Numerical values expressed in terms of PTV and RMS percentage of $|O_R(\mathbf{s})|$ are provided in Table 3. Of particular interest is the image of Figure 8b, where are observed a series of replications of the original object O#3 arranged at the nodes of a square regular grid. This cloning effect is a direct consequence of the convolution product (3) between the object angular brightness distribution O(**s**) and the function $PSF_m(\mathbf{s})$ [27-28]. The point is further discussed in the next paragraph.

Table 3: Difference between reconstructed sky objects $O_R(\mathbf{s})$ expressed in terms of PTV and RMS percentage of the maximal modulus.

| Sky maps | Peak to Valley (PTV) | Root Mean Square (RMS) | 2D or 3D plots |
|---|---|---|---|
| O#1 | 1.62 % | 0.33 % | Not shown |
| O#2 | 2.87 % | 0.57 % | Not shown |
| O#3 | 1.14 % | 0.25 % | Figure 8d |

According to the previous results, one may conclude that both types of interferometers have equivalent effective FoV. The latter shall not be infinite however, because it is subject to another limitation, only depending on the geometrical arrangement of the input pupil array. Here two different cases may be distinguished:

1) If the input telescope array is highly redundant (as in the here above considered case of 49 telescopes ordered on a regular square grid), the useful FoV shall be limited by the presence of the multiple side lobes surrounding the original sky object (as shown in Figure 8b). In order to avoid overlap between adjacent images, the FoV should ultimately be limited by the Nyquist criterion implying that its multiplicative product with the angular resolution (itself related to the maximal entrance baseline B) should not exceed the total number N of telescopes [28]. In no way a remapped output pupil concept associated with the phase-shifting technique could remove this fundamental constraint.

2) Non-redundant telescope arrays such as proposed by Golay [29] may actually be more favorable in terms of useful FoV since the previous limitation does not hold. Nevertheless the dynamic range of the system may be altered since the spurious images are known to turn into a diffuse background in that case [28].

In conclusion, it must be emphasized that the ultimate FoV limit should be related to the sole geometry of the input telescope array, and that the proposed technique is applicable to any type of redundant or non-redundant configuration.

*Signal to Noise Ratio*

Signal to Noise Ratio (SNR) achievable by multi-aperture interferometers has also been the scope of numerous publications (see for example the early Refs. [30-31]). Herein is only considered one approximate expression for the fringes visibility where the background noise, in particular, has been neglected. Hence using the here employed mathematical symbols:

$$\text{SNR}_n \approx |\hat{O}(P_n)| N_P \tau \Big/ \big[|\hat{O}(P_n)| N_P \tau + n_P \text{RON}^2\big]^{1/2} \qquad (16)$$

where additional parameters $N_P$, $\tau$ and RON respectively stand for the total number of photons collected by the whole interferometer array, the detector integration time for one single exposure and the read-out noise of the detector array. The previous expression reduces to $SNR_n \approx [|\hat{O}(P_n)|N_P \tau]^{1/2}$ when photon noise dominates (high flux levels, bright objects), or to $SNR_n \approx |\hat{O}(P_n)|N_P \tau / (n_P^{1/2} RON)$ otherwise (low flux levels, faint objects). These simplified formulas show that:

1) For faint objects the SNR is proportional to $n_P^{-1/2}$, which shows the superiority of the linearly remapped output pupil interferometer with respect to the Fizeau type, because the total number of pixels $n_P$ can be reduced dramatically: in the presented simulations for example, $n_P$ was reduced by a factor $75/729 \approx 0.1$ along the Y"-axis, which should translate into a SNR gain around 3. Although this gain may seem modest, it could be significantly improved by the choice of a higher anamorphic ratio, since the phase extraction method only requires a few pixels along the vertical axis.

2) For both faint and bright objects, the above expressions should be multiplied by a factor $\sqrt{M}$ due to the combination of the multiple phase-shifted exposures (Eq. 10). However this apparent advantage is obviously counterbalanced by the global acquisition time being multiplied by a factor M.

Moreover, it should be noted that the redundancy of the output sub-pupil array also influences the global SNR as demonstrated by Roddier [32], who showed that redundant configurations are better suited to faint sky objects (and conversely non-redundant arrays are more favorable for brighter targets). Additionally the phase-shifting technique may also improve the global radiometric efficiency and SNR, because it allows operating with regularly spaced output sub-pupils as sketched in Figure 5, which may be packed into a much smaller area along the Y'-axis than feasible in non-redundant combination schemes. Hence fewer pixels are required in the image plane with the double benefit of higher irradiance levels and lower RON. As an example a classical 2-6-5-4-3 combining arrangement for six telescopes [10] could be reduced by a factor 2 (i.e. simply becoming 2-2-2-2-2), resulting in a SNR gain roughly equal to $\sqrt{2}$, a figure that is intuitively felt to increase with the total number N of collecting telescopes. This extra radiometric advantage remains however to be confirmed by means of additional numerical simulations, but it may already be concluded that the presented interferometer design should be especially efficient for the observation of faint and extended astrophysical objects.

*Chromatism*

Since the fringe patterns generated by the combining optics depicted in Figure 3 are spectrally dispersed along the Y"-axis, it may seem at first sight that each interferogram can be considered as monochromatic and that the proposed design is naturally immune to chromatic effects. This is not absolutely true however, because it was previously assumed in section 2 that the consecutive phase-shifts $\phi_m$ of the reference sub-pupil are introduced by means of its internal delay line. As the addition of an extra OPD $\xi_m$ actually generates a phase step $\phi_m = 2\pi\xi_m/\lambda$ that is wavelength-dependant, this should in practice reduce the useful spectral range noticeably (say, to $\delta\lambda/\lambda = 5\%$ typically [20]). If a more extended wavelength range is required, one may think of incorporating achromatic phase plates into the interferometer arms in order to cancel the $1/\lambda$ dependence. This would evidently involve a more complicated optical design, but the technique is nowadays well known and has been validated by several experimental results achieved in the field of nulling interferometry (see for example Ref. [33]).

## 5.2  Practical considerations

Although the optical components of the interferometer collecting and combining optics have not been described with great detail in section 2, it seems that they are all rather classical (e.g. matrices of tip-tilt mirrors have already been considered for ground or space-based astronomical applications, and anamorphic optics are already employed on some existing ground combiners [9-10]), with the noticeable exception of the pupil remapping optics subsystem. The latter could practically be realized with the help of a few different existing technologies, among which conventional "bulk" optics, monomode optical fiber arrays and integrated optics are briefly discussed below.

*Conventional optics*

A natural idea for designing and manufacturing a pupil remapping subsystem made of conventional optical components (either reflective or dioptric) is to adapt the existing concept of "image slicers" to our own specific need: these systems are known to relay both the FoV and pupil images of the entrance telescope with the required image quality, and to be extremely stable because they are constituted of two or three arrays of mini-mirrors or mini-lenses bonded together by molecular adhesion (also known as optical contacting). In fact the use of such optical components has already been popularized by the success of integral field spectroscopy, and has become common in ground telescope instrumentation: their manufacturing, aligning and testing processes are nowadays well mastered [34-35]. Moreover, it seems that the technique is now mature for space operation.

*Single Mode optical Fibers (SMF)*

A second natural idea is to replace all the intermediate optical components of the relaying optics with single mode optical fibers: SMF present indeed the specific advantage of filtering all the entrance wavefront errors, and could also replace the bulk ODL subsystem as in the experiment described in Ref. [18]. Practically, they could be connected to the individual input and output sub-pupils of the interferometer by means of micro-lenses arrays having different geometries. This technology has already been developed by different teams in order to couple the six telescopes of the CHARA array [10] or to remap the entrance pupil of a ground-based monolithic telescope [36]. It could most probably be adapted to the case of a linearly remapped output pupil interferometer with no great additional effort. However the solution should impose the presence of mini-dioptric focusing components, whose intrinsic chromatism would need to be compensated for by other elements.

*Integrated optics*

One last, ideal solution would be to replace the whole beam collecting optics with one single three-dimensional integrated optics component, which may solve the practical issue of SMF instabilities due to bending and environmental changes. The development of this technique has only started very recently, but its first results are already promising [37]: it consists in locally increasing the refractive index of a glass substrate mounted on a three-translation stage by exposition to the pulses of a femto-second laser. Experiments have shown that three-dimensional monomode channels can effectively be carved into the glass substrate, but the technique still remains to be validated in terms of throughput, purity of the fundamental Gaussian mode, and crosstalk between the different channels.

### 5.3 Calibration and link with phase closure

Sparse aperture interferometers are in practice always affected by instrumental errors corrupting the measurements of MCF amplitudes $ô_n$ and phases $\theta_n$. In this section we only consider the case of phase errors, that may originate either from static OPDs resulting from optical aberrations or manufacturing errors, or from varying OPDs created by the atmospheric seeing above each telescope for ground-based facilities. Compensating for such errors is often achieved with the help of phase closure techniques, which was originally introduced in radio-interferometry [38] and later employed for optical and IR image reconstruction [39-41]. The principle of the calibration method is schematically illustrated in Figure 9.

For the sake of simplicity, let us only consider a triplet of input sub-pupils $P_1$, $P_2$ and $P_3$ (using a central reference sub-pupil is not necessary here). Following classical principles of stellar interferometry, to each pair of telescopes can be associated a given angular frequency and a MCF phase denoted $\theta_{12}$, $\theta_{23}$ or $\theta_{31}$. But each telescope is at the same time affected with random piston errors denoted $\xi_1$, $\xi_2$ and $\xi_3$, and only a phase closure quantity $\Theta_{123} = \theta_{12} + \theta_{23} + \theta_{31}$ can be determined independently from the piston errors. The knowledge of $\Theta_{123}$ can nevertheless be supported by additional measurements performed in a phase-shifting mode: varying the internal OPDs in $P_1$ would for example give access to the following quantities:

$$\Theta_{12} = \theta_{12} + \xi_2$$

$$\Theta_{13} = -\theta_{31} - \xi_3. \tag{18a}$$

Likewise, phase shifting of $P_2$ and $P_3$ sub-apertures with respect to their neighbors provides the following additional information:

$$\Theta_{23} = \theta_{23} + \xi_3$$

$$\Theta_{21} = -\theta_{12} - \xi_1, \text{ and} \tag{18b}$$

$$\Theta_{31} = \theta_{31} + \xi_1$$

$$\Theta_{32} = -\theta_{23} - \xi_2. \tag{18c}$$

Eqs. 18.a-c together with the phase closure quantity $\Theta_{123}$ form a system of linear equations that can be pseudo-inversed in a least-squares sense, finally allowing to retrieve both the object phases $\theta_{12}$, $\theta_{23}$ and $\theta_{31}$ and the instrumental errors $\xi_1$, $\xi_2$ and $\xi_3$. Generalizing this procedure to the whole array would lead in principle to a full calibration of its instrumental errors that could be performed regularly on a spaceborne facility. Moreover, it allows removing $2\pi$ ambiguities that may affect the phase closure quantity $\Theta_{123}$ without employing heavy reconstruction algorithms [42-43]. It is likely, however, that this calibration procedure would be hardly applicable to ground-based interferometers affected by atmospheric disturbance, although an approaching method was studied in more detail by Greenaway [44] in the frame of the study of a large monolithic ground telescope.

# 6   SUMMARY

In this paper was described the principle of a multi-aperture interferometer associated with a phase-shifting technique and suitable for quick, snapshot imaging of astrophysical objects through Fourier inversion. A few advantages of the proposed design were highlighted with the help of numerical simulations, among which are radiometric efficiency, Field of View equivalent to those of Fizeau interferometers, and a preliminary calibration procedure allowing full characterization of the instrumental errors. But the most important of these advantages probably is the considerable simplification of the optical and mechanical design of the relaying and combining optics: this is made possible because the whole interferometer system do not need to satisfy the classical input/output pupils "golden rule" characterizing Fizeau interferometers, while still delivering images in quasi real-time if the total number of telescopes is high enough.

The results of the presented numerical simulations suggest that it may not be unrealistic to coherently couple hundreds of free-flying telescopes on a single 4K x 4K detector chip, allowing a spectral decomposition of the object at the same time, and using only conventional optical components or emerging technologies. The major remaining critical issue should then be to control the numerous spacecrafts flying in formation, and to coarsely phase them within the adjustment range of the on-board delay lines. It may be hoped that future space projects searching for habitable extra-solar planets will contribute to validate this technology, therefore paving the way for the concrete realization of such "indirect imaging arrays" in space.

The author would like to thank the anonymous reviewers for many remarks and suggestions resulting in appreciable clarification of the original manuscript.

**FIGURES CAPTIONS**

Figure 1: General layout of the multi-aperture interferometer array.
Figure 2: Schematic view of a single collecting telescope and its launching optics made of a delay line and tip-tilt mirror.
Figure 3: Schematic layout of combining optics.
Figure 4: Coordinate systems used for the collecting telescopes array (top) and combining optics of the interferometer (bottom).
Figure 5: Geometrical arrangement of input and output interferometer pupils (red circles indicate locations of the reference sub-pupils).
Figure 6: Flow-chart of image simulations and object reconstruction codes.
Figure 7: Major image reconstruction steps of the linearly remapped output pupil interferometer. From left to right and top to bottom: original sky object O#2 (a), its images seen through a 5-m and 1-m diameter telescope (b and c), phase-shifted fringe patterns formed in the image plane O"X"Y" (d and e) and their difference map (f), grey scale maps of the modulus and phase of function $OTF_C(P)$ (g and h) and a 3D view of its modulus (i), reconstructed MCF modulus (j), 3D views of the modulus (k) and imaginary part (l) of the reconstructed sky object $O_R(s)$, and its grey scale map (m).
Figure 8: Difference between images $O_R(s)$ reconstructed by the linearly remapped output pupil and Fizeau interferometers. From left to right and top to bottom: original sky object O#3 (a), its image formed by a Fizeau interferometer (b), grey scale map of reconstructed images (c), and 3D view of their difference map (d). PTV and RMS errors are indicated in Table 3.
Figure 9: Possible calibration procedure applicable to three individual sub-apertures.
1.

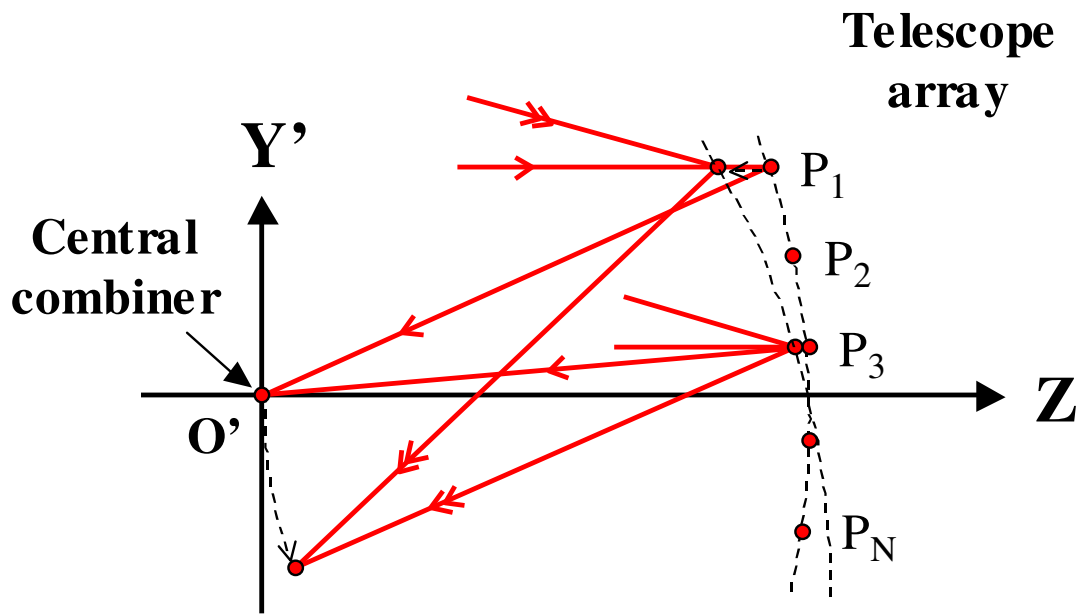

**Figure 1: General layout of the multi-aperture interferometer array.**

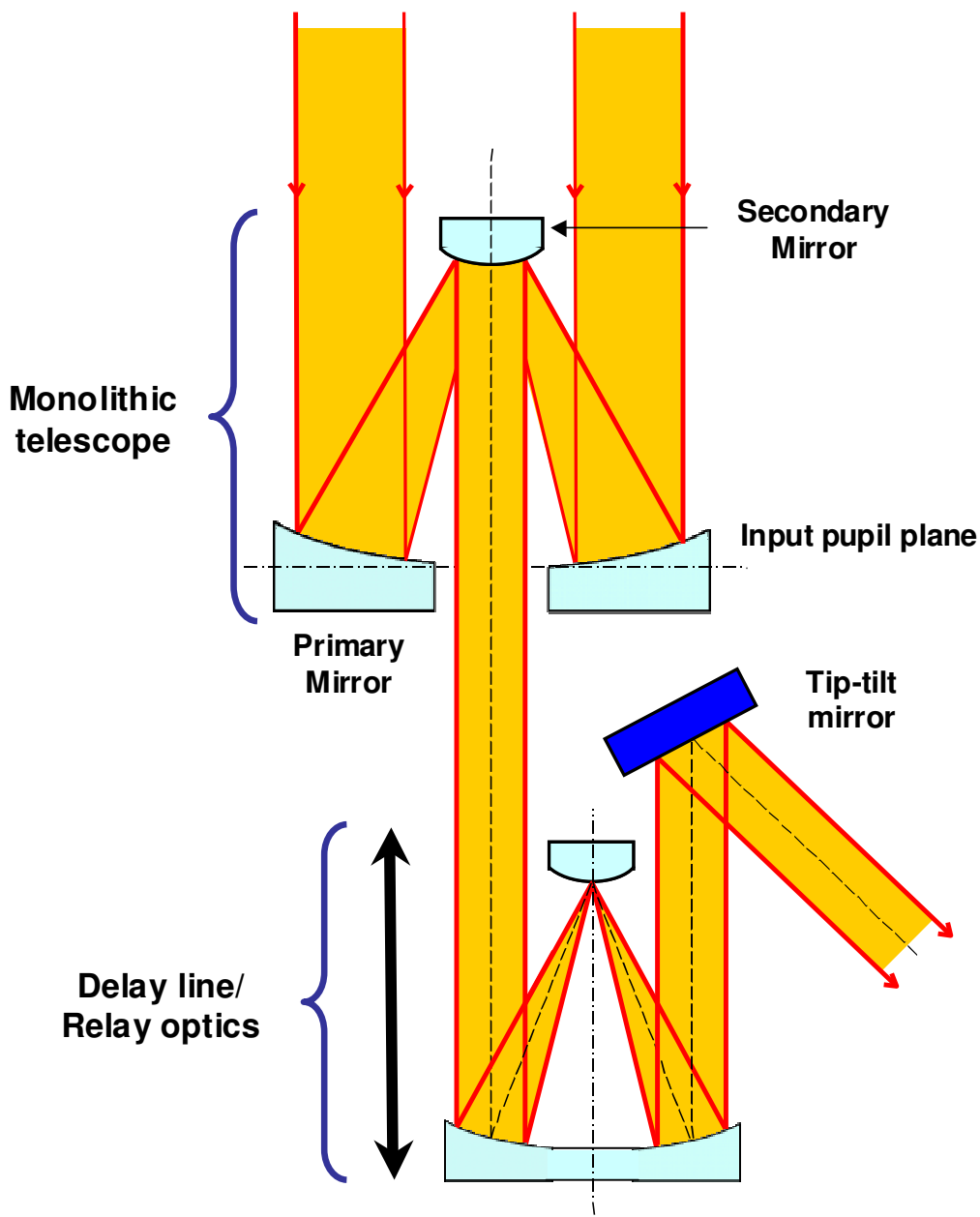

**Figure 2: Schematic view of a single collecting telescope and its launching optics made of a delay line and tip-tilt mirror.**

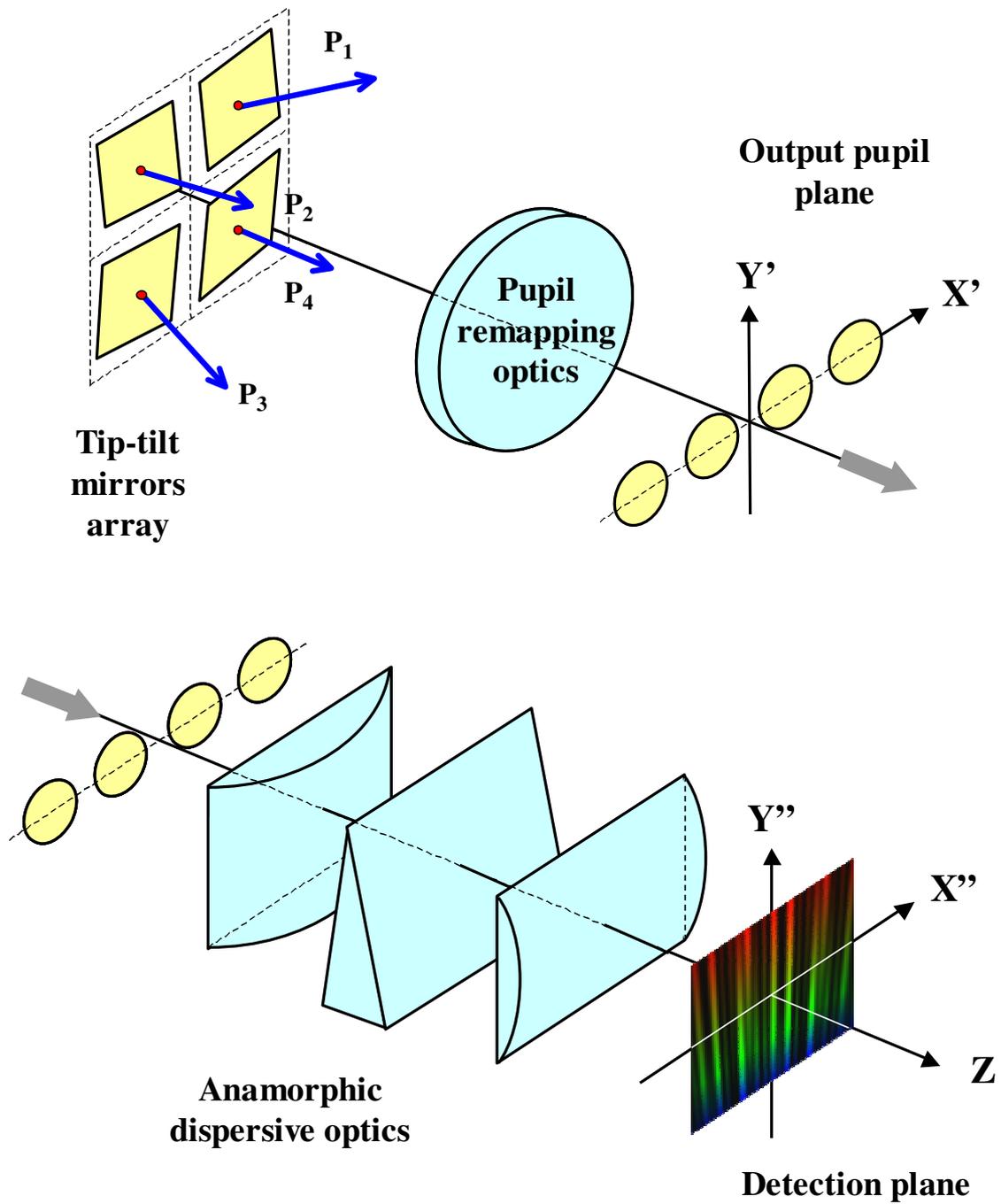

**Figure 3: Schematic layout of combining optics.**

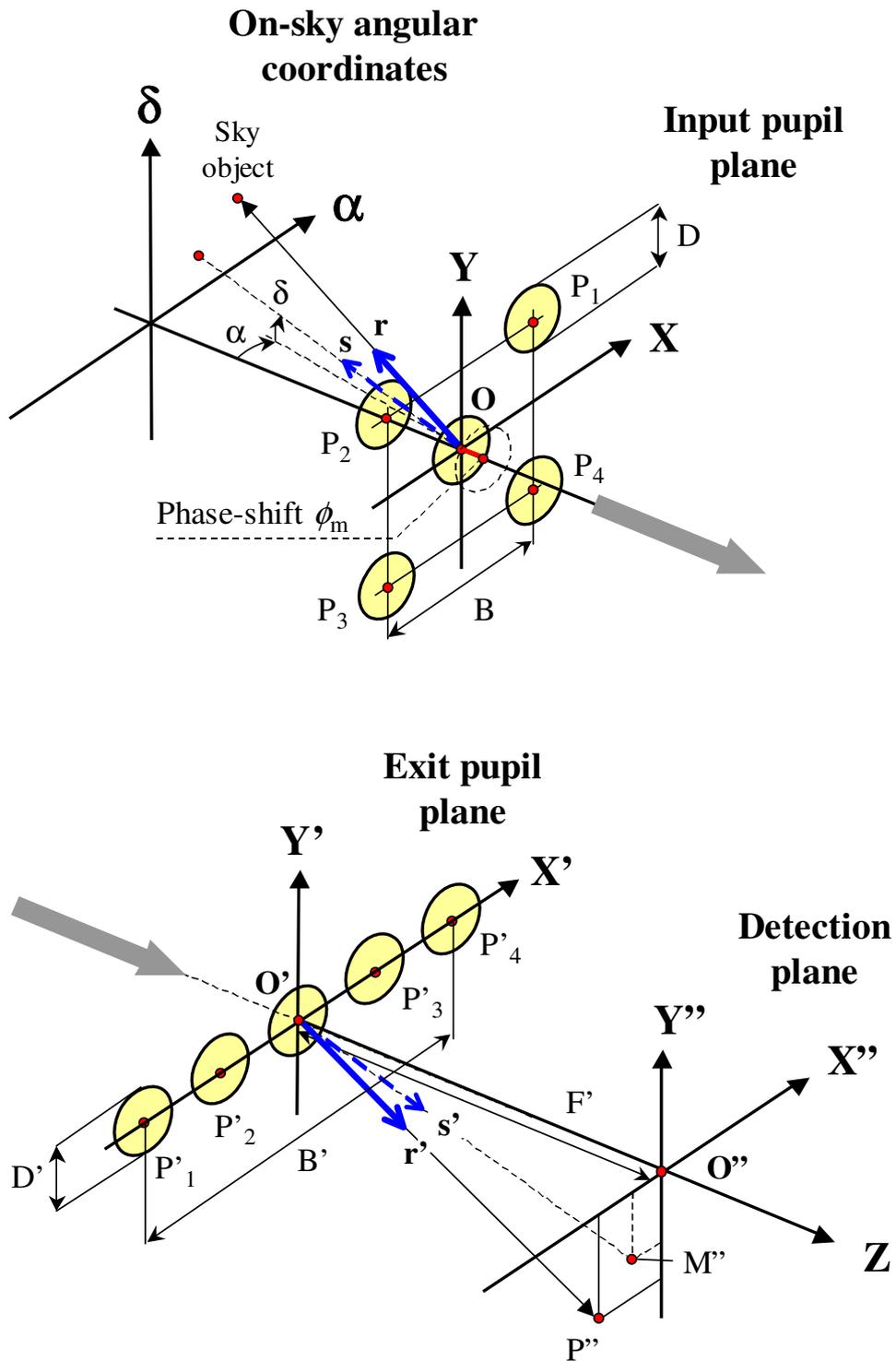

**Figure 4:** Coordinate systems used for the collecting telescopes array (top) and combining optics of the interferometer (bottom).

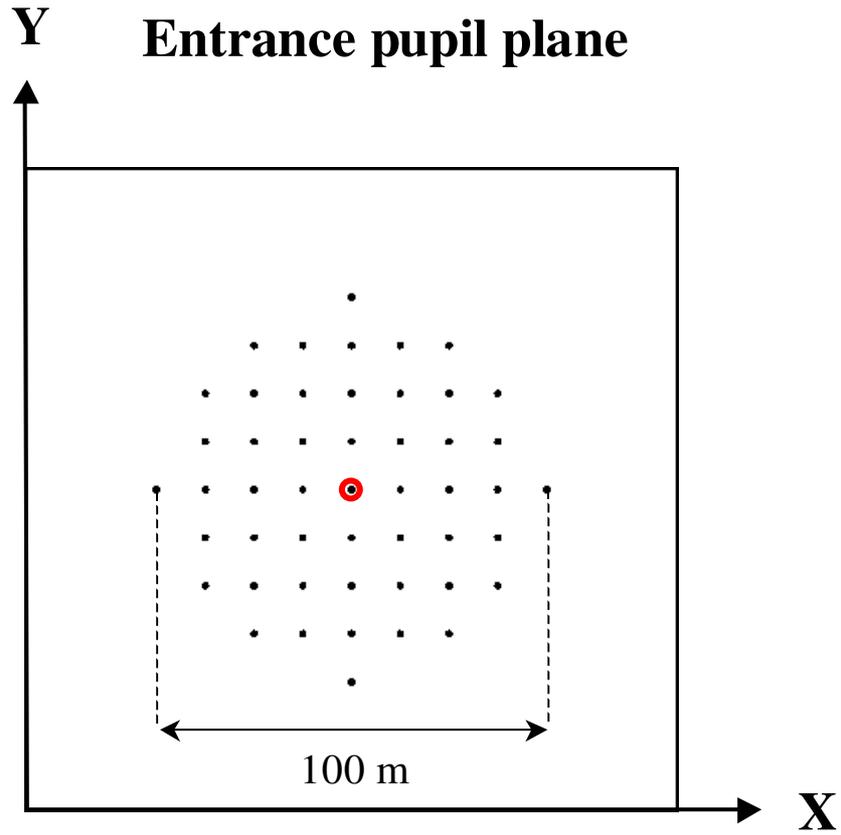

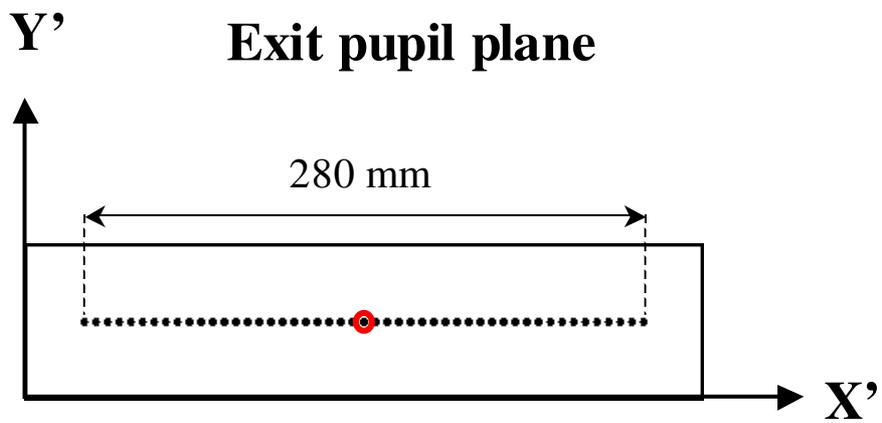

**Figure 5: Geometrical arrangement of input and output interferometer pupils (red circles indicate locations of the reference sub-pupils).**

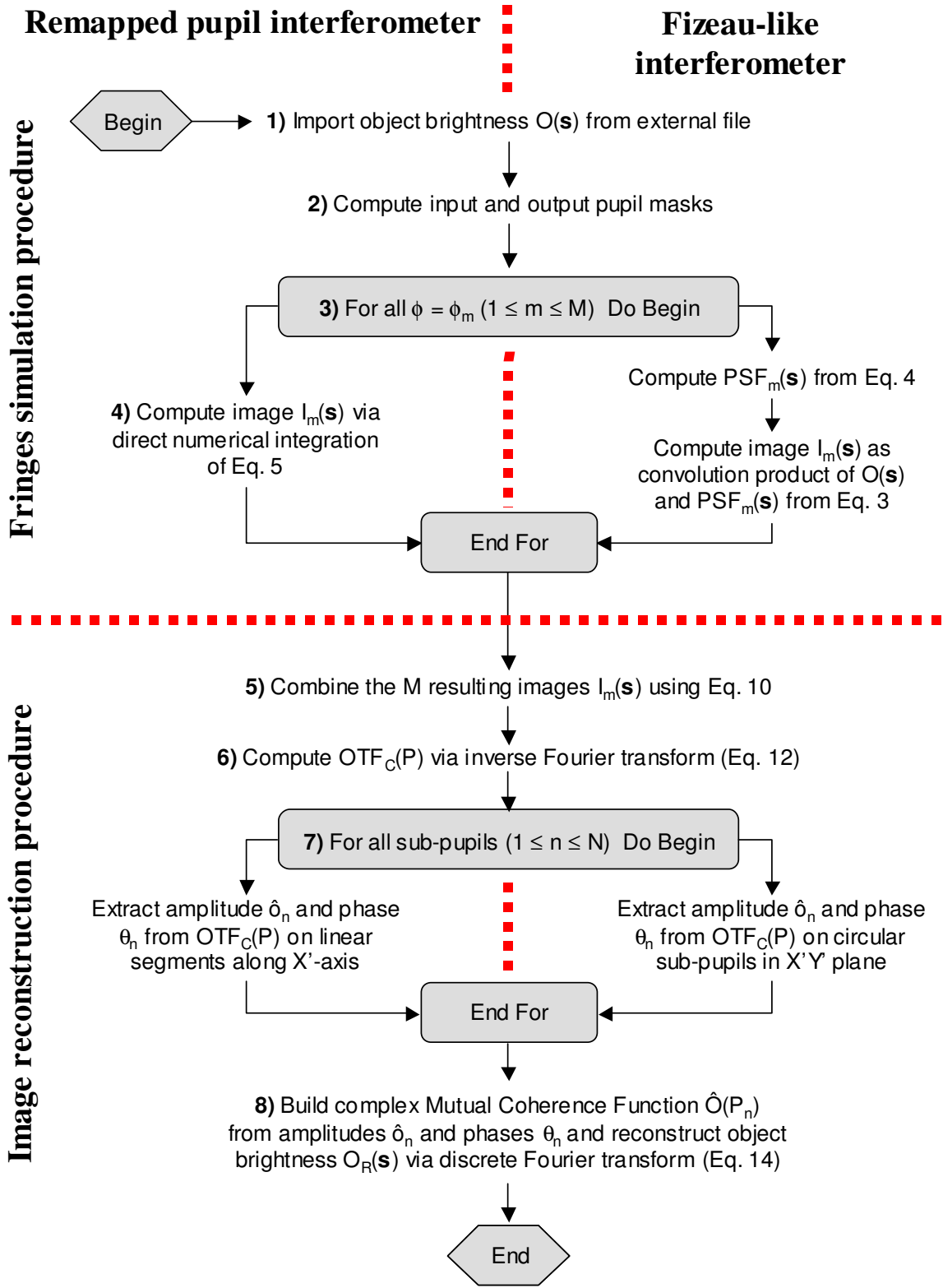

**Figure 6: Flow-chart of image simulations and object reconstruction codes.**

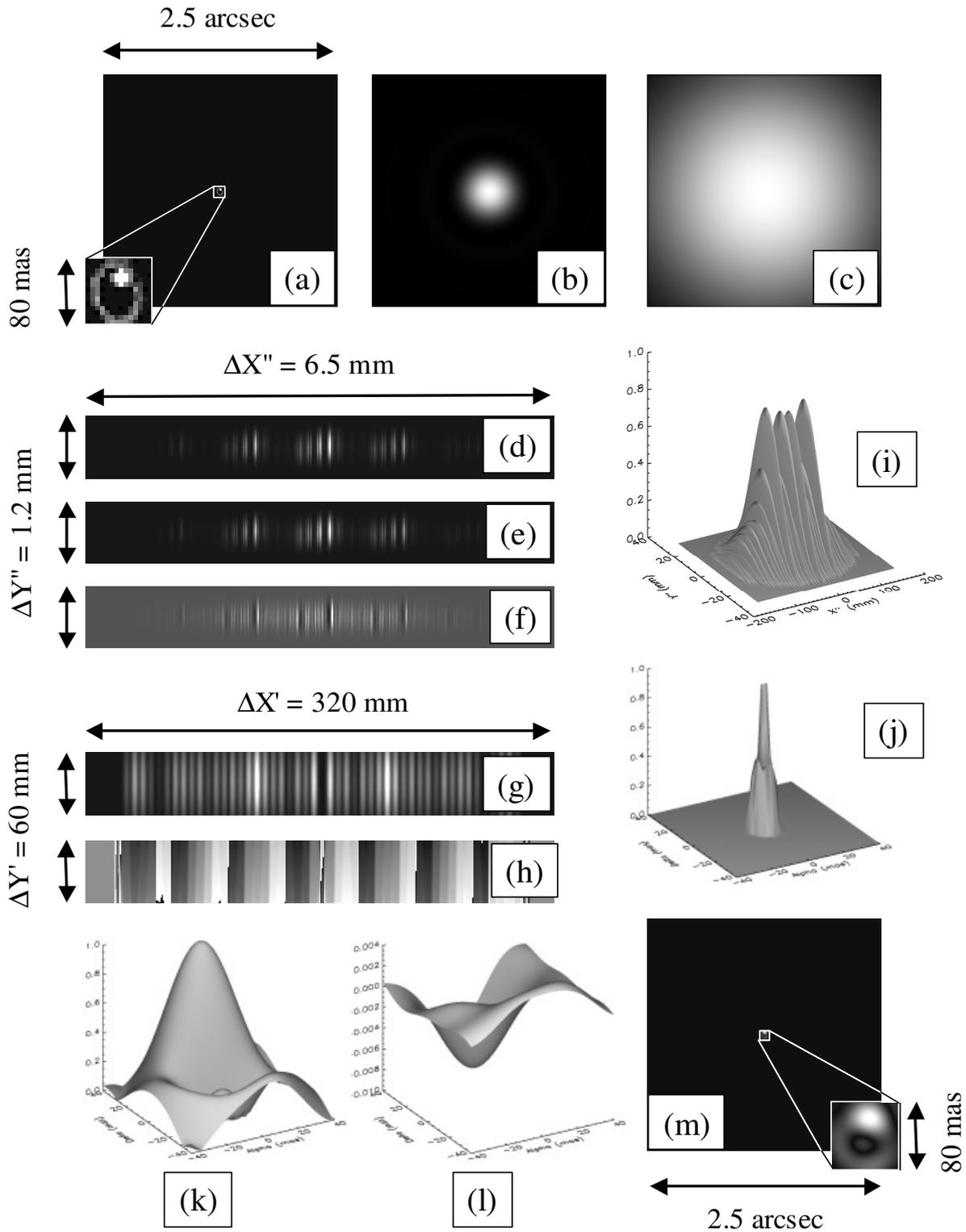

**Figure 7: Major image reconstruction steps of the linearly remapped output pupil interferometer. From left to right and top to bottom: original sky object O#2 (a), its images seen through a 5-m and 1-m diameter telescope (b and c), phase-shifted fringe patterns formed in the image plane O"X"Y" (d and e) and their difference map (f), grey scale maps of the modulus and phase of function $OTF_C(P)$ (g and h) and a 3D view of its modulus (i), reconstructed MCF modulus (j), 3D views of the modulus (k) and imaginary part (l) of the reconstructed sky object $O_R(s)$, and its grey scale map (m).**

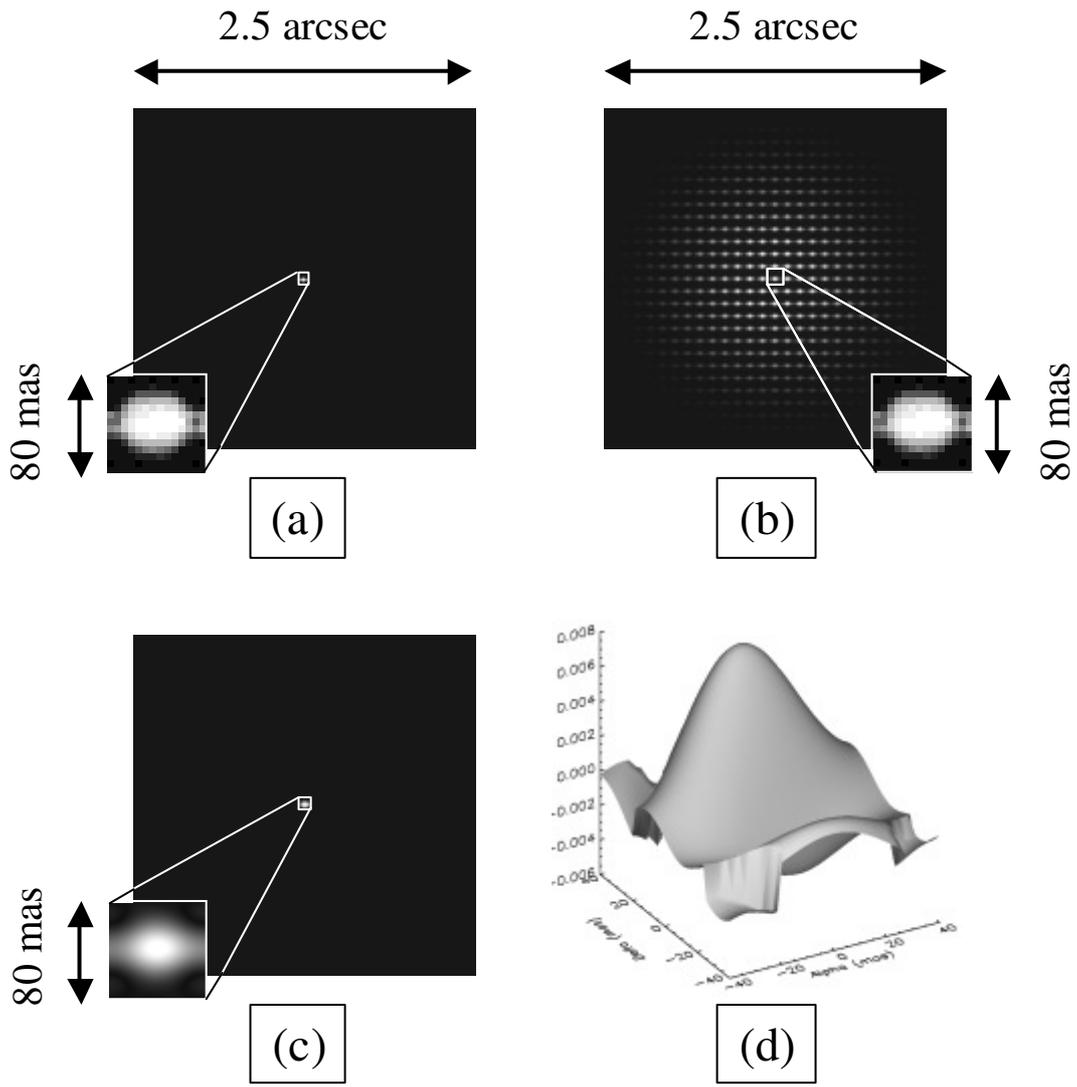

**Figure 8:** Difference between images $O_R(s)$ reconstructed by the linearly remapped output pupil and Fizeau interferometers. From left to right and top to bottom: original sky object O#3 (a), its image formed by a Fizeau interferometer (b), grey scale map of reconstructed images (c), and 3D view of their difference map (d). PTV and RMS errors are indicated in Table 3.

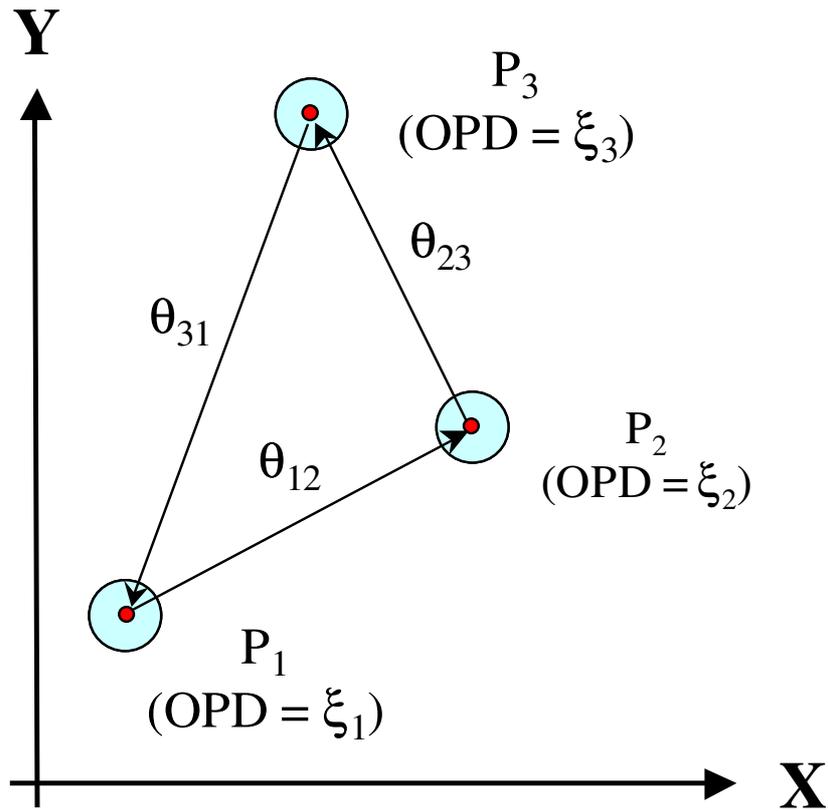

**Figure 9: Possible calibration procedure applicable to three individual sub-apertures.**